\documentclass [aps,pre,twocolumn,amsmath,amssymb,showpacs]{revtex4}

\usepackage[dvips]{graphicx}
\usepackage{psfrag}

\DeclareMathOperator{\sech}{sech}

\begin{document}

\title{ Soliton Ratchets Induced by Excitation of Internal Modes} 

\author{C.R. Willis}
\author{M. Farzaneh}
\affiliation{ Department of Physics, Boston University, 590 Commonwealth Ave,
 Boston, MA, 02215}
\date{\today}
\begin{abstract}
Recently Flach et.al.\cite{Flach02} used a symmetry analysis to
predict the appearance of directed energy current in homogeneously
spatially extended systems coupled to a heat bath in the presence of
an external ac field $E(t)$. Their symmetry analysis allowed them to
make the right choice of $E(t)$ so as to obtain symmetry breaking
which causes directed energy transport for systems with a nonzero
topological charge. Their numerical simulations verified the existence
of the directed energy current. They argued that the origin of their
strong rectification in the underdamped limit is due to the excitation
of internal modes and their interaction with the translational kink
motion. The internal mode mechanism as a cause of current
rectification was also proposed by Salerno et.al.\cite{Salerno02}. We
use a rigorous collective variable for nonlinear Klein-Gordon
equations to prove the rectification of the current is due to the
excitation of an internal mode, $\Gamma(t)$, which describes the
oscillation of the slope of the kink, and to a dressing of the bare
kink by the ac driver. The internal mode $\Gamma(t)$ is excited by its
interaction with the center of mass of the kink, $X(t)$, which is
accelerated by $E(t)$. The external field $E(t)$ also causes the kink
to be dressed. We derive the expressions for the dressing and
numerically solve the equations of motion for $\Gamma(t)$, $X(t)$, and
the momentum, $P(t)$, which enable us to obtain the explicit
expressions for the directed energy current and the ac driven kink
profile. We then show the directed energy current vanishes unless the
slope $\Gamma(t)$ is a dynamical variable and the kink is dressed by
the ac driver.
\end{abstract}
\maketitle

\section{INTRODUCTION}

In a recent paper S. Flach et.al.\cite{Flach02} studied the appearance
of directed energy currents in homogeneous spatially extended systems
described by nonlinear field equations coupled to a heat bath in the
presence of an external ac field $E(t)$. As pointed out in
\cite{Flach02}, rectifying energy transform using fluctuations has
been studied in connection with such problems as molecular motors in
biological systems \cite{Julicher97}, electrical currents in superlattices
\cite{Seeger78,Goychuk98,Alekseev98,Alekseev}, voltages in Josephson
junction coupled systems \cite{Zapata98,Weiss00,Reimann02} and other problems.

The authors of Ref.[1] showed by a symmetry analysis that the correct
choices of $E(t)$ lead to directed energy transport for nonlinear
Klein-Gordon systems with a nonzero topological charge. They used
numerical simulations of the ac driven Klein-Gordon equation which
confirmed their predictions which generalized recent rigorous theories
of currents generated by broken time-space symmetries to the case of
interacting many particle systems \cite{Flach00,Yevtushenko01}. They
did this by replacing the fluctuations as a superposition of ac
driving fields and uncorrelated white noise. They also showed the
persistence of directed currents in the Hamiltonian limit of systems
exposed to ac fields but decoupled from the heat bath. The authors of
Ref.[1] then argued that the origin of the observed strong
rectification in the underdamped limit is due to the nonadiabatic
excitation of internal kink modes and their interaction with the
translational kink motion.

In this paper we use a rigorous collective variable (CV) theory for
nonlinear Klein Gordon equations derived in
Refs.[13,14]\nocite{Boesch98,Boesch90} to prove that an external ac
field causes the CV's for the center of mass $X(t)$ and the slope
$\Gamma(t)$ to become time dependent and to interact with each
other. The ac driver in addition to inducing time dependence in $X(t)$
the center of mass of the kink and the slope $\Gamma(t)$ causes the kink to be
dressed by phonons. The dressing changes the shape of the kink and
increases the coupling of $\Gamma(t)$ to $X(t)$. We show the non-vanishing of the
dressing is a necessary condition for breaking time inversion symmetry
of the energy current. However, the dressing of the kink alone in the
absence of a time dependent collective variable, $\Gamma(t)$, cannot cause
current rectification. Consequently, we prove that the existence of
the time dependence of $\Gamma(t)$ and the dressing $\chi(t)$ are necessary for time
inversion symmetry breaking.

In Sec.II we derive the equations of motion for $X(t)$ and
$\Gamma(t)$ including the terms due to the dressing of the kink by
phonons. We present our results for the solutions $X(t)$ and
$\Gamma(t)$ and for the generation of directed energy currents in
Sec.III and in Sec.IV and we discuss our results. The derivation of
the dressing $\chi$ is given in the appendix.

\section{DERIVATION OF CV EQUATIONS OF MOTION}

Before deriving the CV equations of motion used in this paper, we will
make a few remarks about CV treatments of the Klein-Gordon
equations. The first approach which is derived in Refs.[13,14] and used
in this paper, is to treat the center of mass $X(t)$ and the slope
$\Gamma(t)$ as collective variables which satisfy coupled second order
differential equations, which also depend on the dressing of the
kink. In this approach the equations of motion for $X(t)$ and
$\Gamma(t)$ are not manifestly relativistic invariant. What has been
proven is that when the solutions $X(t)$ and $\Gamma(t)$ are inserted
in the kink $\phi[X(t), \Gamma(t)]$ that $\phi$ satisfies the
relativistic invariant nonlinear Klein-Gordon equation. An analogous
well-known example of a non-manifestly relativistic case is the use of
the non-relativistic Coulomb gauge which leads to the relativistic
solution $E(t)$ and $B(t)$ of Maxwell's equations. The second approach
would be to consider a single CV, $X(t)$. The equation of motion for
$X(t)$ is fourth order in time because the Lagrangian contains the
second derivative $\ddot{X}(t)$. The Lagrangian of a theory that
contains a second derivative, $\ddot{X}(t)$, leads to equations of
motion that contain the fourth derivative $\ddddot{X}(t)$. The fourth
order equation for $X(t)$ has the same number of degrees of freedom as
the equivalent CV theory for two variables which consists of two coupled second
order equations for $X(t)$ and $\Gamma(t)$.

We outline the derivation of the equations of motion for the
collective variables $X(t)$ and $\Gamma(t)$ which are derived in
detail in Refs.[13,14]. The damped nonlinear SG equation for the field
$\phi$ in the presence of an external potential, $V(\phi)$, is
\begin{equation}
 \phi_{,tt} - \phi_{,xx} + \sin\phi + \beta \phi_{,t} =
   -\frac{\partial V}{\partial\phi} 
\end{equation}
where $\beta\phi_{,t}$ is the damping due to the heat bath, and where
we are using dimensionless variables where the velocity of the phonons
is $c =1$. We introduce the collective variables by writing the
solution $\phi$ of Eq.(1) in the form
\begin{equation}
 \phi(x,t) = \sigma [\,\xi(t)\,] + \chi [\,\xi(t)\,]
\end{equation}
where $\xi\equiv\Gamma(t)[\,x-X(t)\,]$ and the single kink solution
$\sigma[\,\xi(t)\,]$ is
\begin{equation}
 \sigma[\,\xi(t)\,] = 4 \tan^{-1}\exp[\,\Gamma(t)\,(\,x-X(t)\,)\,]
\end{equation} 
and $\chi[\,\xi(t)\,]$ is the dressing of the kink by phonons due to the
external potential $V(\phi)$ which for the applied ac field of this
paper is given by
\begin{equation}
 V(\phi) = \bigg(\,\epsilon_1\cos\omega t + \epsilon_2\cos[\,2\omega t +
            \theta\,]\,\bigg)\,\phi(\,\xi(t)\,)\equiv f_1\,\phi(\,\xi(t)\,). 
\end{equation}
$\theta$ is an arbitrary phase and $\epsilon_1$, and $\epsilon_2$ are
perturbation parameters i.e., we solve for $\phi$ to first order in
$\epsilon_1$ and $\epsilon_2$. The CV's are the center of mass $X(t)$
and the slope of the kink evaluated at its center is $2\,\Gamma(t)$.
In Ref.[2] \nocite{Salerno02} a directed kink motion for the SG was
obtained numerically for the first time using $V(\phi)$ in Eq.(4) for
a wide range of momenta with an analytic approach for small momenta.

The equations of motion for $X$ and $\Gamma$ each contain many terms
proportional to integrals of $\chi$, its time derivatives and spatial
derivatives. $\chi$ is a solution of the linearized ac driven SG
equation and is proportional to $\epsilon_1$ and $\epsilon_2$. In the
appendix we solve for $\chi$. The solution for $\chi$ is
\begin{equation}
 \chi = \frac{4}{\pi}\, f(t)\, \sech^2 \xi
\end{equation}
where
\begin{eqnarray*}
f(t) &=&(\epsilon_1/2)\cos\omega t
        \,\bigg\{\frac{1-\omega}{\beta^2+(1-\omega)^2}+
              \frac{1+\omega}{\beta^2+(1+\omega)^2}\bigg\} \\
     &+&(\epsilon_2/2)\cos(2\omega t + \theta)
         \,\bigg\{\frac{1-2\omega}{\beta^2+(1-2\omega)^2} \\
     & &\hspace{1.3in}+\frac{1+2\omega}{\beta^2+(1+2\omega)^2}\bigg\}
        {\rm (\ref{ap7})}
\end{eqnarray*}
Since $\chi$ is an even function of $\xi$ many of the terms in
Eqs.(2.4a) and (2.5a) of Ref.[13,14] for $\ddot{X}$ and $\ddot{\Gamma}$
that depend on integrals of $\chi$ vanish. The only terms which
survive are
\begin{eqnarray}
 (1-b_X) M_X \bigg[\ddot{X} + \dot{X}(\dot{\Gamma}/\Gamma)+\beta
             \dot{X}\bigg]& = & 2\pi f_1 \nonumber \\
             &+&\Gamma^2\langle\sigma'|\chi''\rangle(1-\dot{X}^2)\nonumber \\
             &-&(\dot{\Gamma}/\Gamma)^2\langle\sigma'|\xi^2\chi''\rangle
             \nonumber \\
             &-&2(\dot{\Gamma}/\Gamma)(\dot{f}/f)\langle\sigma'|\xi\chi\rangle
             \nonumber \\
             &-&(\ddot{\Gamma}/\Gamma)\langle\sigma'|\xi\chi\rangle
\end{eqnarray}
where $M_X\equiv \Gamma\langle\sigma'|\sigma'\rangle = 8\Gamma$ ,
$b_X\equiv(\Gamma/M_X)\langle\sigma''| \chi\rangle = 0$ and where
$\langle f|g\rangle \equiv \int f^*(\xi)\,g(\xi)\,d\xi$.
The corresponding equation for $\ddot{\Gamma}$ is
\begin{widetext}
\begin{equation}
 (1-b_\Gamma) M_\Gamma \bigg[\ddot{\Gamma} -3 \dot{\Gamma}^2/2\Gamma
             +(M_X/2\Gamma)(1-\dot{X}^2)+\beta
             \dot{\Gamma}\bigg]=
             (2\dot{X}\dot{\Gamma}/\Gamma^2)\langle\xi\sigma'|\chi''\xi\rangle
             +
             2(\dot{X}/\Gamma)(\dot{f}/f)\langle\xi\sigma'|\chi'\rangle
             + (\ddot{X}/\Gamma)\langle\xi\sigma'|\chi'\rangle
\end{equation}
\end{widetext}
where $M_\Gamma\equiv \Gamma^{-3}\langle\xi\sigma'|\xi\sigma'\rangle =
(2\pi^2/3\Gamma^3)$ and
             $b_\Gamma\equiv(\Gamma^3/M_\Gamma)^{-1}\langle\xi^2\sigma''|
             \chi\rangle = 0$ .

We next eliminate the $\ddot{\Gamma}$ term in Eq.(6) and the
$\ddot{X}$ term in Eq.(7) by using the zeroth order in $\epsilon_1$,
and $\epsilon_2$ equations for $\ddot{X}$ and $\ddot{\Gamma}$. The
elimination is justified because the corresponding terms are both
multiplied by $\chi$ which is already first order in $\epsilon_1$, and
$\epsilon_2$. 

The zeroth order expression for $\ddot{X}$ is
$\ddot{X}=-\dot{X}(\dot{\Gamma}/\Gamma)$ and for $\ddot{\Gamma}$ is
$\ddot{\Gamma} = 3\dot{\Gamma}^2/2\Gamma - M_X ( 2\Gamma
M_\Gamma)^{-1}(1-\dot{X}^2)$.
When we substitute for $\ddot{X}$ in Eq.(7), we obtain
\begin{eqnarray}
 \ddot{\Gamma}+\beta\dot{\Gamma}&=&(3\dot{\Gamma}^2/2\Gamma)+(6/\pi^2)\Gamma^3(1-\dot{X}^2)
  \nonumber \\
  &+&(3/2\pi^2)\bigg[8\dot{\Gamma}\dot{X}\big(2\langle\xi\sigma'|\chi''\rangle\nonumber
  \\
  & &\hspace{0.5in}-\langle\xi\sigma'|\chi'\rangle\big)
  +8\dot{f}\Gamma^2\dot{X}\langle\xi\sigma'|\chi'\rangle\bigg]
\end{eqnarray}
and when we substitute for $\ddot{\Gamma}$ in Eq.(6) we obtain
\begin{eqnarray}
 M_X \bigg[\ddot{X} + \dot{X}(\dot{\Gamma}/\Gamma)+
 \beta\dot{X}\bigg]= 2\pi f_1
-2\dot{f}(\dot{\Gamma}/\Gamma)\langle\sigma'\xi|\chi'\rangle & &
 \nonumber \\
+\Gamma^2(1-\dot{X}^2)\bigg(\langle\sigma'|\chi''\rangle-
(6/\pi^2)\langle\sigma'|\xi\chi'\rangle\bigg)& & \nonumber \\
-(\dot{\Gamma}/\Gamma)^2\bigg(\langle\sigma'|\xi^2\chi'\rangle-3/2\langle\sigma'\xi|\chi'\rangle
 \bigg)
\end{eqnarray}
The momentum $P$ conjugate to $X$ is $
P=M_X\dot{X}=8\Gamma\dot{X}$. Consequently we can write Eq.(8) for
$\ddot{X}$ in terms of $P$ i.e,
\begin{eqnarray}
\frac{dP}{dt}+\beta P =2\pi
f_1-2\dot{f}(\dot{\Gamma}/\Gamma)\langle\sigma'\xi|\chi'\rangle& &
\nonumber \\
+\big(\Gamma^2-\frac{P^2}{64}\big)\bigg(\langle\sigma'|\chi''\rangle
-\frac{6}{\pi^2}\langle\sigma'|\xi\chi'\rangle \bigg)& & \nonumber \\
-(\frac{\dot{\Gamma}}{\Gamma})^2\bigg(\langle\sigma'|\xi^2\chi''\rangle-3/2
\langle\sigma'\xi|\chi'\rangle\bigg)
\end{eqnarray}
and Eq.(8) for $\ddot{\Gamma}$ in terms of $P$ becomes
\begin{eqnarray}
 \ddot{\Gamma}-(3\dot{\Gamma}^2/2\Gamma)-(6/\pi^2)\Gamma\bigg[1-\Gamma^2+(P/8)^2\bigg]
 + \beta\dot{\Gamma} & & \nonumber \\   
 =(3/2\pi^2)\bigg[P\dot{\Gamma}\bigg(2\langle\xi\sigma'|\chi''\rangle
 -\langle\xi\sigma'|\chi'\rangle\bigg) & & \nonumber \\
 +\Gamma P (\dot{f}/f)\langle\xi\sigma'|\chi'\rangle\bigg]
\end{eqnarray}
Finally after evaluating the integrals and replacing $\dot{X}^2$ by
$P^2/64$ we obtain the final form of our equations of motion for
$P(t)$ and $\Gamma(t)$
\begin{eqnarray}
 \dot{P}+\beta P -2\pi f_1 =& f(t)&\bigg[0.47(8/\pi)(\dot{\Gamma}/\Gamma)^2 \nonumber \\
 &-&(8/3)\bigg(\dot{f}(t)/f(t)\bigg)
 (\dot{\Gamma}/\Gamma) \nonumber \\
 &-&(8/\pi^2)\bigg(\Gamma^2-(P/8)^2\bigg)\bigg]
\end{eqnarray}
and
\begin{eqnarray}
\ddot{\Gamma}+\beta\dot{\Gamma}-3\dot{\Gamma}^2/2\Gamma-(6/\pi^2)\,\Gamma\,\bigg(1-\Gamma^2+(P/8)^2\bigg)=
 & & \nonumber \\
  f(t)(2\pi^2)^{-1}\bigg[\bigg(5/2-\pi^2/16\bigg)\,P
 \dot{\Gamma}-\bigg(\dot{f}(t)/f(t)\bigg)\,P\Gamma\bigg]
\end{eqnarray}

Before solving the equations of motion for $P$ and $\Gamma$ it is
worth making a few remarks about the properties of the coupled
equations. In Eq.(12) the ac driver, $f_1(t)$, directly drives
$\dot{P}$ while in Eq.(13) for $\ddot{\Gamma}$ the ac driver,
$f_1(t)$, does not directly drive $\ddot{\Gamma}$ because the $\Gamma$
mode,$\frac{\partial \sigma}{\partial \Gamma}$, is orthogonal to
the ac driver. However, both $\dot{P}$ and $\ddot{\Gamma}$ see the ac
driver indirectly through the dressing $\chi$ which is proportional to
$f(t)$, Eq.(A.7) which also depends on the two frequencies $\omega$
and $2\omega$. As long as $\omega<0.5$ the phonon radiation is
small. In this paper we consider only frequencies which are much less
than one, which is the beginning of the lower band edge in the units
of this paper.  Consequently in this paper the emission of phonons is
negligible. Since there are no modes of $X(t)$ and $\Gamma(t)$ in the
band gap of the SG, there is no excitation of internal gap modes by
the ac driver as there is e.g., in $\phi^4$ and the double
sine-Gordon. The dressing $\chi$ which is proportional to $\epsilon$
changes the shape modes to $\sigma_{,X}+\chi_{,X}$ and to
$\sigma_{,\Gamma}+\chi_{,\Gamma}$ in addition to changing the
frequency of $P$ and $\Gamma$ directly.

Before discussing the results we give a discussion of the symmetries
of the coupled Eqs.(12) and (13). The first symmetry is refered to as
the shift symmetry of the driver which is $P \rightarrow -P$ and $t
\rightarrow t + \frac{\tau}{2}$ provided $f_1(t) =
-f_1(t+\frac{\tau}{2})$ and $f(t) = -f(t+\frac{\tau}{2})$ are always
shift symmetric if and only if a Fourier expansion contains only odd
terms. Thus $f_1(t)$ and $f(t)$ in Eqs.(12) and (13) always violate
shift symmetry. A second symmetry is time inversion symmetry i.e., $P
\rightarrow -P$ when $t \rightarrow -t$ and $\beta = 0$. Eqs.(12) and
(13) satisfy time inversion symmetry when $\beta = 0$ and $\theta =
0,\pm n\pi$. When $\beta$ is small, time inversion is approximately
satisfied.
  
\section{RESULTS OF SIMULATIONS}

In this section we present the computer solutions of Eqs.(12) and (13)
for $\Gamma(t)$, $P(t)$, and for the time average of the energy
current. In our units the energy current $J(t)$ is equal to $P(t)$
because
$$
 J(t) \equiv -\int_{-\infty}^\infty \sigma_{,t}(\xi)\,\sigma_{,x}(\xi)dx
 = 8\,\Gamma\,\dot{X}=P(t)
$$
where $\sigma[\,\xi\,]=4\tan^{-1}\exp[\,\Gamma(t)\big(x-X(t)\big)\,]$.  Consequently
the time average of $J(t)$, $\langle J(t) \rangle$, is equivalent to the time
average of $P(t)$, $\langle P(t) \rangle$. 

In Fig.(1) we show the results for $\langle P(t) \rangle$ for a range
of values $\omega$, $\beta$, $\epsilon_1$, and $\epsilon_2$ which show
clearly the directed energy current as a function of $\theta$. We see
that there is symmetry breaking for all the sets of parameter values.
\begin{figure}
  \psfrag{x1}{$\theta$}
  \psfrag{x2}{$\langle P \rangle$}
  \includegraphics*[width=3in]{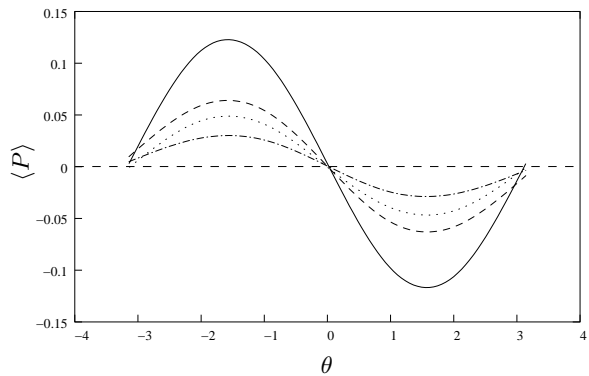}
  \caption{This figure demonstrates symmetry breaking,i.e., the
           nonvanishing of $\langle P(t) \rangle$ as a function of $\theta$ for 
           various values of the parameters $\omega$, $\epsilon$ and
           $\beta$. Solid curve: $\omega=0.1$, $\epsilon_1=\epsilon_2=
           0.03$ and $\beta=0$ ( The curve is multiplied by 0.15),
           dashed curve: $\omega=0.3$, $\epsilon_1=0.3$, $\epsilon_2=
           \epsilon_1/\sqrt 3$ and $\beta=0.2$, dotted curve:
           $\omega=0.1$, $\epsilon_1=\epsilon_2=0.05$, and
           $\beta=0.12$, dash-dotted curve: $\omega=0.25$,
           $\epsilon_1=0.16$, $\epsilon_2=\epsilon_1/\sqrt 2$ and
           $\beta=0.15$. $\langle P(t) \rangle$ has the units of
           momentum and $\theta$ is in radians.}
\end{figure}
\begin{figure}
  \psfrag{x1}{$\theta$}
  \psfrag{x2}{$\langle P \rangle$}
  \includegraphics*[width=3in]{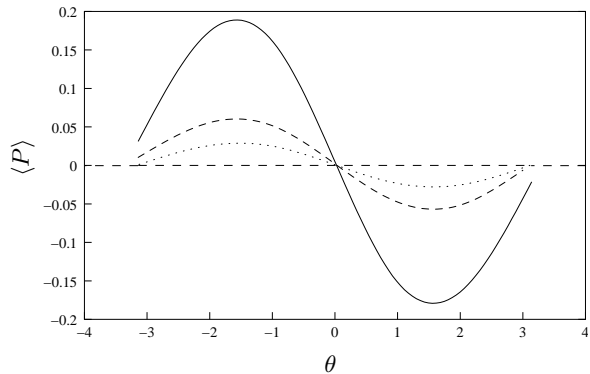}
  \caption{$\langle P(t) \rangle$ as a function of $\theta$ for
           $\omega=0.1$, $\epsilon_1=\epsilon_2=0.03$ for various values of
           $\beta$ show a monotonic decrease of the amplitude of $\langle P(t)
           \rangle$ as the damping $\beta$ increases. Solid curve: $\beta=0.02$,
           dashed curve: $\beta=0.05$, dotted curve: $\beta=0.12$.  In this
           simulation $\langle P(t) \rangle$ decreases as $\beta$ increases but
           the values of $\langle P(t) \rangle$ are so small that they can not be
           distinguished on the figure. $\langle P(t) \rangle$ has the units of
           momentum and $\theta$ is in radians.}
\end{figure}
In Fig.(2) we show  $\langle P(t) \rangle$ as a function of $\theta$
for fixed values of $\omega$ and $\epsilon_1 = \epsilon_2$ and various
values of $\beta$. If $\beta\neq 0$ then time inversion symmetry is
not valid. However as $\beta$ goes to zero, time inversion symmetry is
approximately restored at $\theta = 0, \pm n\pi$. Thus we observe for 
small $\beta$ as $\beta$ decreases, exactly the same behaviour as in
Fig.(1) of Ref.[1] that is the smaller $\beta$ the larger the value of  
$\langle P(\theta = 0) \rangle$, but the smaller the value of $\theta$
at which $\langle P(\theta) \rangle = 0$. The values in Fig.(2) at 
$\theta = 0$ are  $\langle P(t) \rangle$ = 0.0008 when $\beta = 0.12$,
$\langle P(t) \rangle$ = 0.002 when $\beta = 0.05$ and $\langle P(t)
\rangle$ = 0.005 when $\beta = 0.02$.

In the three typical examples of $P(t)$ and $\Gamma(t)$, Figs.(3), (4)
and (5) we have selected the shapes and magnitudes vary
considerably. The shapes of the $P(t)$ and $\Gamma(t)$ curves show the
effect of being driven by an ac driver with two frequencies which
causes the curves to vary in amplitude and shape when we vary the
parameters $\omega$, $\epsilon_1$, $\epsilon_2$, and $\beta$. The
changes in shape of $\Gamma(t)$ are often striking because $\Gamma(t)$
is a very nonlinear oscillator which has a complicated response to the
ac driver and the dressing $\chi$. Whereas in lowest order, the
equation for $P(t)$ is linear. Generally the magnitudes of both
$\Gamma(t)$ and $P(t)$ increase with increases in the strength of
$\epsilon_1$, and $\epsilon_2$. The variable $\Gamma(t)$ oscillates
about an average value of $\langle\Gamma(t) \rangle$ which is greater
than $\Gamma_0=1$, the unperturbed kink value of $\Gamma$. It usually
also takes instantaneous values less than one. The relative change in
slope, $\Delta\Gamma/\Gamma$, varies from a few percent to as much as
one hundred percent and is strongly dependent on the magnitude of
$\epsilon_1$ and $\epsilon_2$. Large values of $\Delta\Gamma/\Gamma$
represent large distortions of the shape of the kink.
\begin{figure*}
\centering 
\begin{tabular}{c c}
  \begin{minipage}{2.5in}
  \psfrag{x1}{$t$}
  \psfrag{x2}{$P(t)$}
  \includegraphics*[width=2.5in]{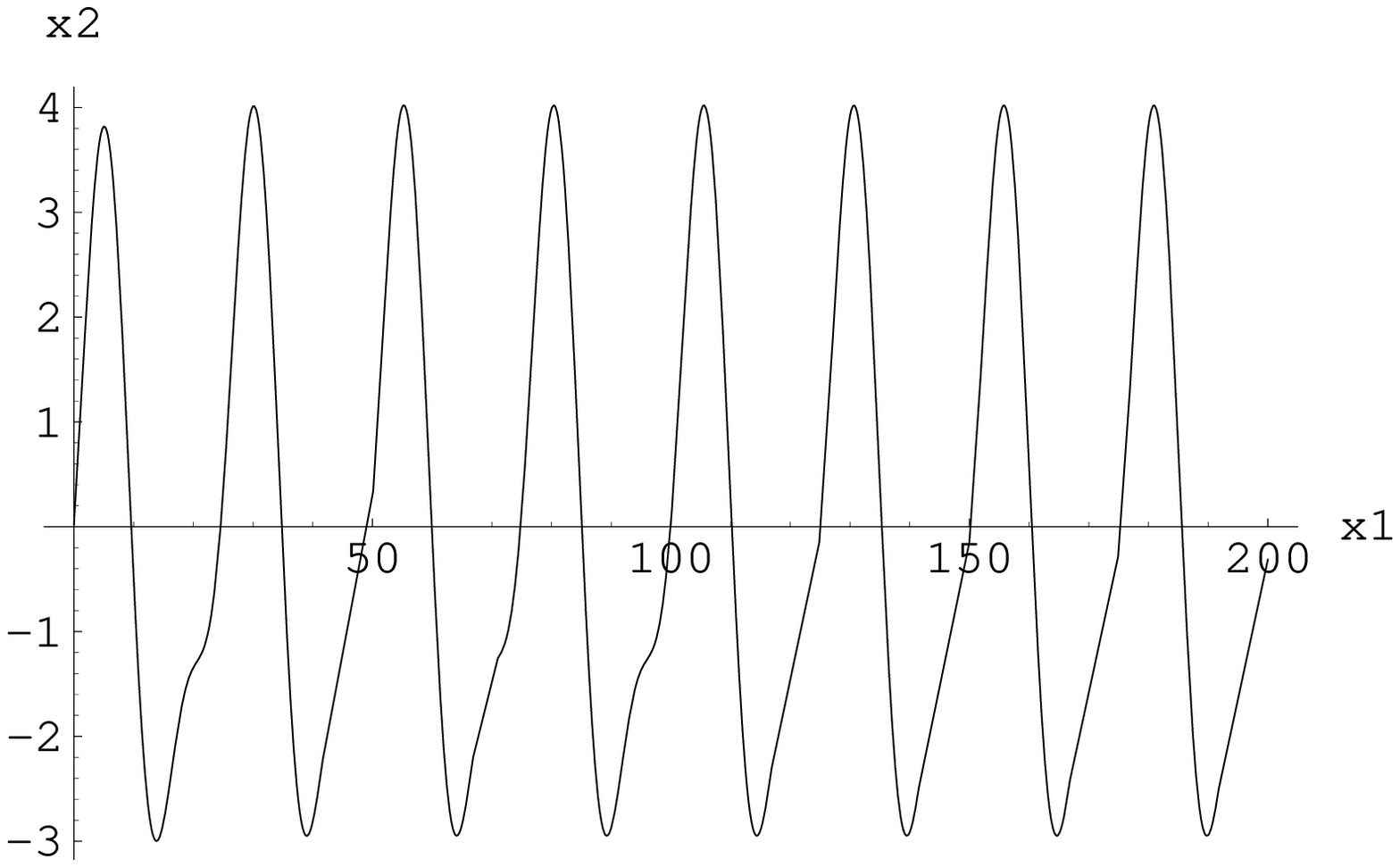}
 \end{minipage} &
 \begin{minipage}{2.5in}
  \psfrag{x1}{$t$}
  \psfrag{x2}{$\Gamma(t)$}
  \includegraphics*[width=2.5in]{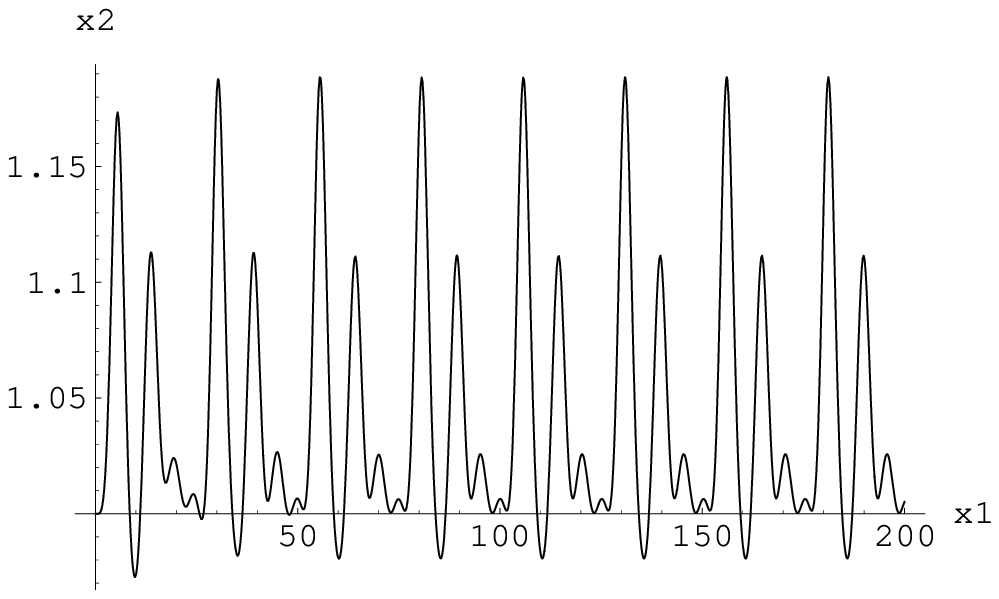}
 \end{minipage}
\end{tabular}
\caption{The energy current $P(t)$ and the slope of the
                       kink $\Gamma(t)$ for the parameters $\omega=0.25$, 
                       $\beta=0.15$, $\epsilon_1=0.16$,
                       $\epsilon_2=\epsilon_1/\sqrt{2}$ and
                       $\theta=1.61-\pi$. Both curves show the effect
                       of two driving frequencies. The time average of
                       $P(t)$ is nonzero and $\Gamma(t)$ has a
                       multiple frequency oscillation about
                       $\langle\Gamma(t) \rangle$ greater than $\Gamma_0$ with an
                       amplitude change
                       $\Delta\Gamma/\langle\Gamma\rangle\sim
                       20\%$. $P(t)$ has the units of momentum and $\Gamma(t)$ has the units
                       of inverse length.}
\end{figure*}
\begin{figure*}
\centering
\begin{tabular}{c c}
  \begin{minipage}{2.5in}
  \psfrag{x1}{$t$}
  \psfrag{x2}{$P(t)$}
  \includegraphics*[width=2.5in]{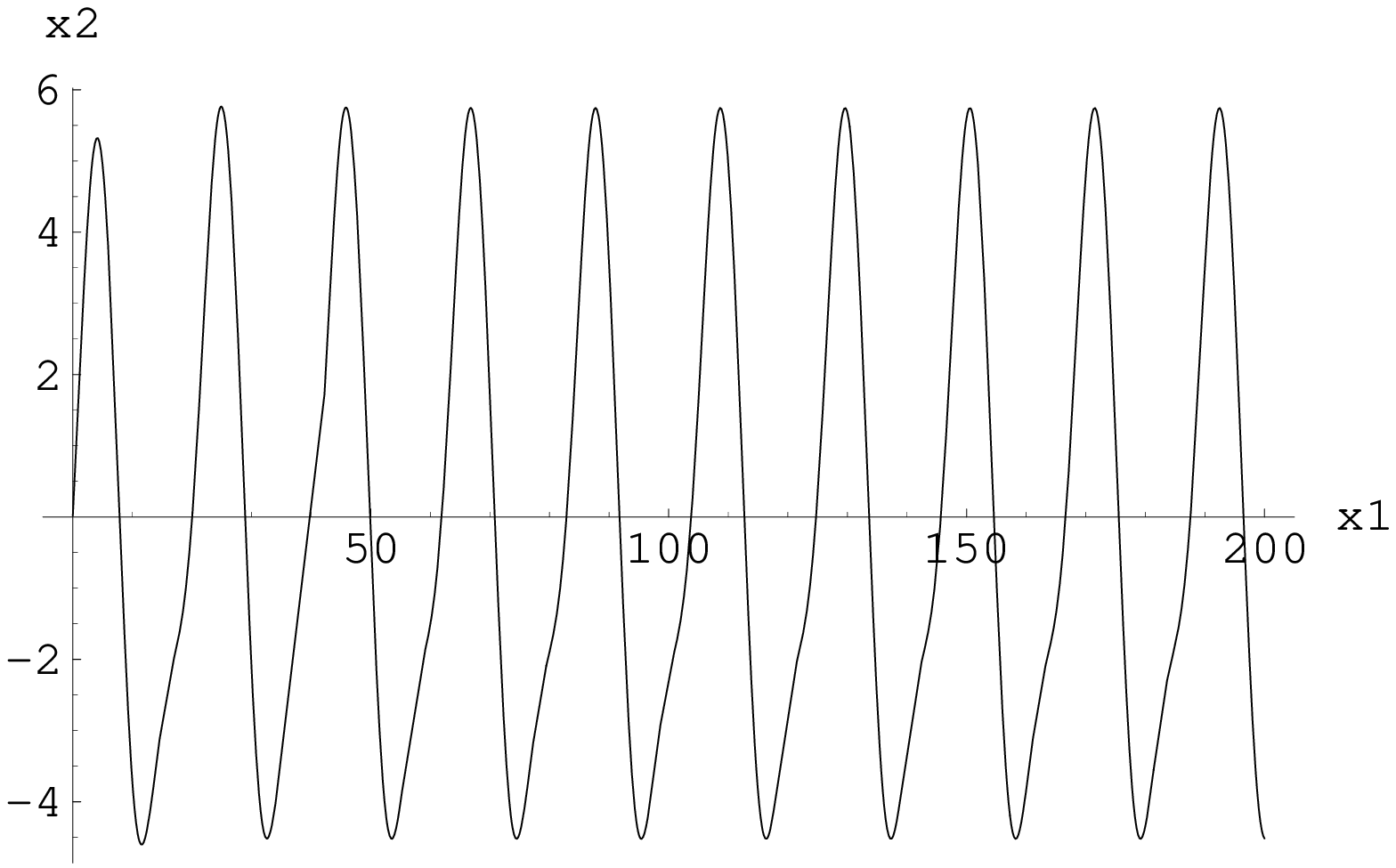}
 \end{minipage} &
 \begin{minipage}{2.5in}
  \psfrag{x1}{$t$}
  \psfrag{x2}{$\Gamma(t)$}
  \includegraphics*[width=2.5in]{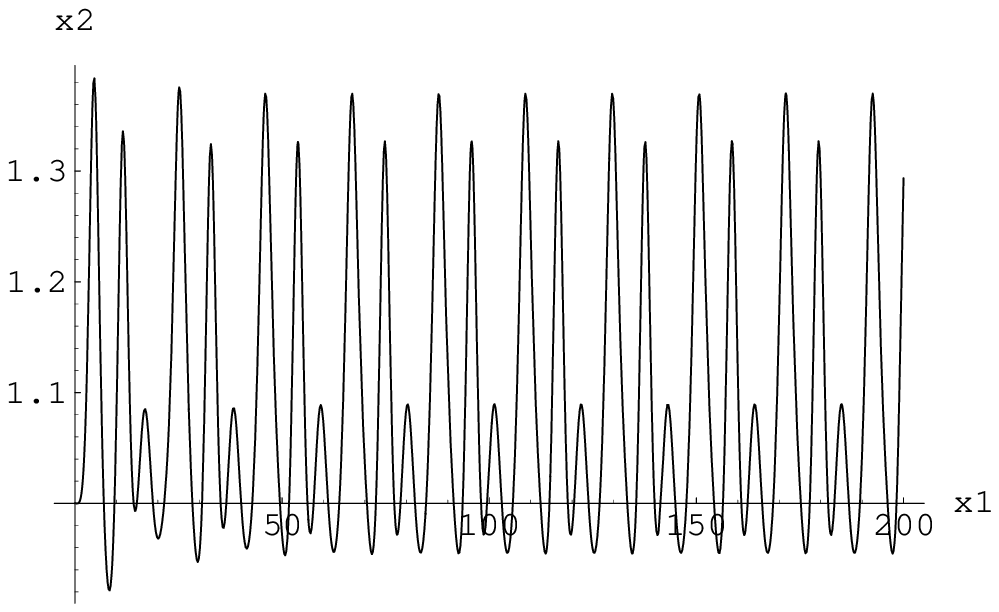}
 \end{minipage} 
\end{tabular}
\caption{$P(t)$ and $\Gamma(t)$ for the parameters
                       $\omega=0.3$, $\beta=0.2$,
                       $\epsilon_1=0.3$, $\epsilon_2=\epsilon_1/\sqrt
                       3$ and $\theta=1.61-\pi$. $\langle P(t) \rangle$ is nonzero
                       and the slope $\Gamma(t)$ has a multiple
                       frequency oscillation about $\langle\Gamma(t) \rangle >
                       \Gamma_0$ with an amplitude change
                       $\Delta\Gamma/\langle\Gamma \rangle \sim 50\%$.
                       $P(t)$ has the units of momentum and $\Gamma(t)$ has the units
                       of inverse length.}
\end{figure*}
\begin{figure*}
\centering
\begin{tabular}{c c}
  \begin{minipage}{2.5in}
  \psfrag{x1}{$t$}
  \psfrag{x2}{$P(t)$}
  \includegraphics*[width=2.5in]{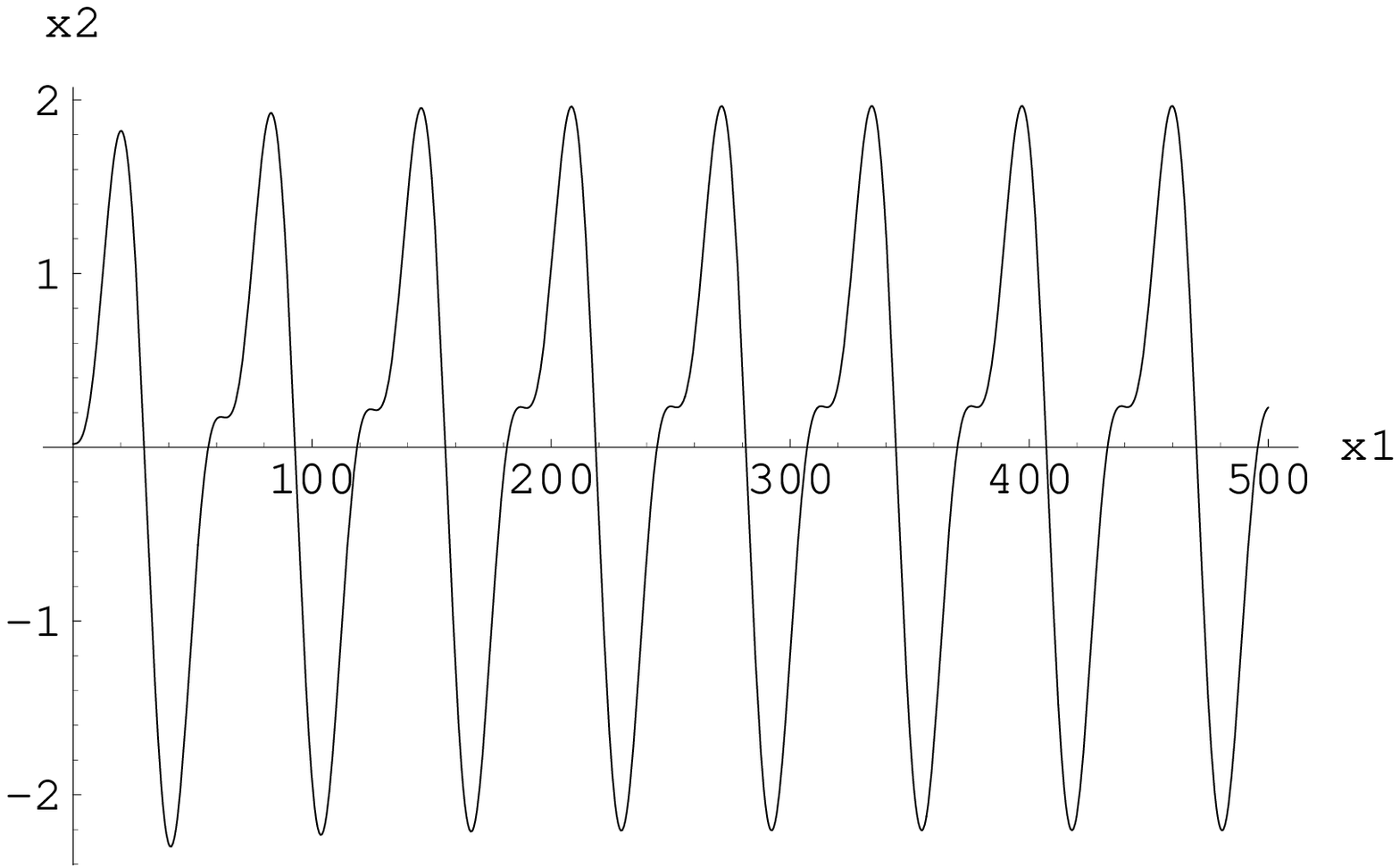}
 \end{minipage} &
 \begin{minipage}{2.5in}
  \psfrag{x1}{$t$}
  \psfrag{x2}{$\Gamma(t)$}
  \includegraphics*[width=2.5in]{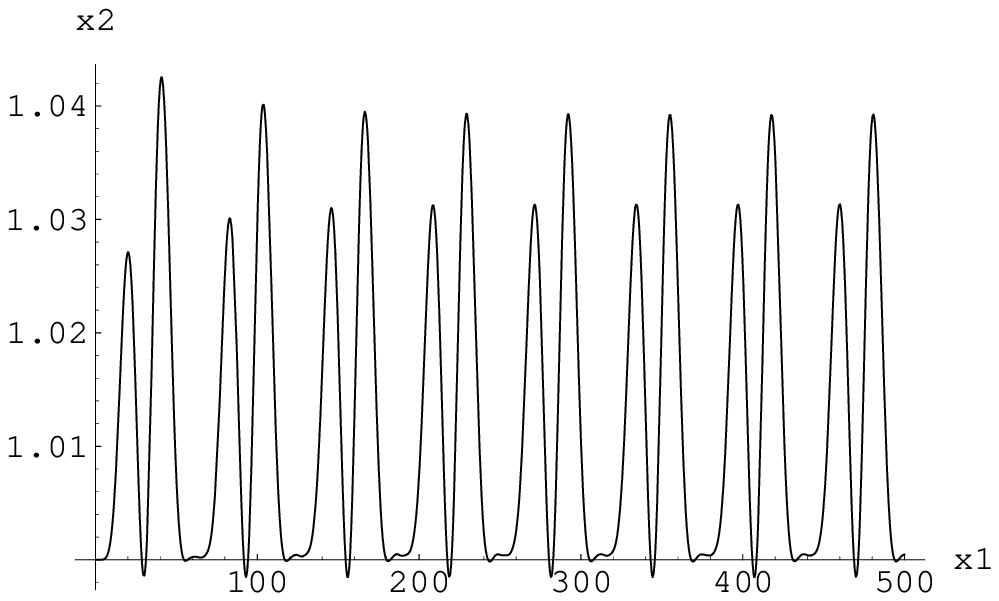}
 \end{minipage}
\end{tabular}
\caption{$P(t)$ and $\Gamma(t)$ for the parameters $\omega=0.1$, 
                       $\beta=0.02$, $\epsilon_1=\epsilon_2=0.03$,
                       and $\theta=\pi$. $\langle P(t) \rangle=0$ as is required by
                       $\theta=\pi$. $\Gamma(t)$ has a two frequency
                       oscillation about $\langle\Gamma \rangle > \Gamma_0$ with
                       only an amplitude change of $\Delta\Gamma /
                       \langle\Gamma \rangle \sim 4\%$. The relatively weak
                       response of $P(t)$ and $\Gamma(t)$ is due to
                       the smallness of the driver $\epsilon$. $P(t)$
                       has the units of momentum and $\Gamma(t)$ has the units
                       of inverse length.}
\end{figure*}
When we compare $P(t)$ and $\Gamma(t)$ for the kink dressed by $\chi$
with the bare kink we find a strong dependence on the phase $\theta$
which leads to different shapes and amplitudes of $P(t)$ and
$\Gamma(t)$ for different $\theta$. For example in Fig.(5) for $\omega=0.1$,
$\beta=0.02$, $\epsilon_1=\epsilon_2=0.03$ and $\theta$ near $\theta=
0, \pm n\pi$ the slope $\Gamma(t)$ of the dressed kink is a pattern of
single peaks while for the bare kink with the same parameters
$\Gamma(t)$ is a pattern of double kinks. On the other hand for
$\theta$ appreciably different from $\theta= 0,\pm n\pi$ e.g., $\theta
= 1.61-\pi$ the differences in shape of $P(t)$ between the bare and
dressed kinks are relatively minor. At the same time although the
shapes of $\Gamma(t)$ are qualitatively similar the bare kinks have
appreciably reduced amplitudes of $\Gamma(t)$.

\section{DISCUSSION}
The ac driver causes the center of mass $X$ and the slope $\Gamma$ to
become time dependent and the kink to be dressed by phonons given by
the expression $\chi(t) = (4/\pi) f(t) \sech^2\xi(t)$. The dressing
$\chi$ which is not a CV internal mode can be observed as a modulation
of the structure of the kink $\sigma$. We proved that the existence of
a directed energy current arises from the existence of the internal
degree of freedom, $\Gamma(t)$, combined with the dressing $\chi(t)$.  We
found that the directed energy current vanished when $\Gamma$ was set
equal to $\Gamma_0$. When we set $\chi =0$ in Eq.(10) for $\dot{P}$
the right hand side vanishes and we obtain
$$
 \dot{P} + \beta P = 2 \pi f_1
$$
The infinite time average of this equation vanishes when we use the
fact that the thermal average of the initial value of $P$
vanishes. Thus there is no directed current in the SG unless the slope
depends on $t$ and the kink is dressed by $\chi$. 

We observe in the computation of $\langle P(t) \rangle$ that the kink
sees the heat bath only through the damping term $ -\beta \langle P(t)
\rangle$ i.e., $\langle P(t) \rangle$ does not see the fluctuations of
the heat bath. The reason is that when we represent the bath as a
generalized Fokker -- Planck equation and calculate $\langle P(t)
\rangle$ the damping term contributes $-\beta \langle P(t) \rangle$
because the damping is represented by $\beta\frac{\partial}{\partial
P}$. However the fluctuation term is proportional to a second
derivative $\partial^2/\partial P \partial P$ and thus gives a
vanishing contribution to $\langle P(t) \rangle$. Note a fluctuation
such as $\langle P^2 \rangle$ or $\langle P(t)P \rangle$ would see
both the damping term and the bath fluctuations.

In conclusion, we have proven that the symmetry breaking that leads to
a directed energy current in the ac driven SG is generated by the
existence of the time dependence of the slope, $\Gamma(t)$, and by the
dressing, $\chi(t)$. In Ref.\cite{Salerno02a}, Salerno and
Quintero showed that a double SG showed ratchet behaviour. In
Ref.\cite{Marchesoni96}, Marchesoni obtained a directed kink
transport by the $\sin\phi$ potential for $\epsilon_2=0$. Costantini
{\em et al.} \cite{Costantini02} observed ratchet behavior in ac driven
assymetric kinks.

\acknowledgments{We would like to thank Sergej Flach for useful
discussions on symmetry breaking.}

\appendix*
\section{}

We calculate the dressing $\chi$ generated by the ac driven $f_1(t)$
in the linear approximation i.e., the lowest order in $\epsilon_1$,
$\epsilon_2$ and by using the Green's function of the linearized SG
equation in the presence of a soliton Eq.(C5) of Ref.[14]. The formal
solution for $\chi$ is:
\begin{eqnarray}
 \chi(\xi)&=&2 \,{\rm Re}\int_{-\infty}^{\infty}dk\,
 \bigg(\Omega(k)\bigg)^{-1}\psi_k(\xi)\int_{-\infty}^{\infty}
 d\xi'\,\psi_k^*(\xi')\nonumber \\
 & &\int_0^\infty
 dt'\sin\Omega(k)(t-t')e^{-\beta(t-t')}f_1(t')
\end{eqnarray}
where $\Omega(k)\equiv (1+k^2)^{1/2}$ and the eigenfunctions of the
linearized SG are
\begin{equation}
 \psi_k(\xi)=(2\pi)^{-1/2}e^{ik\xi}\bigg[ik-\tanh\xi\bigg]
\end{equation}
The $\xi'$ integral is: 
\begin{eqnarray}
 \int_{-\infty}^{\infty}d\xi'\,\psi_k(\xi')&=&-ik\int_{-\infty}^{\infty}d\xi'\cos
 k\xi'\nonumber \\
 & -& \int_{-\infty}^{\infty}d\xi'\sin k\xi'\tanh\xi' 
\end{eqnarray}
The first integral is an irrelevant constant which we can neglect. 
Integrating the second integral by parts we obtain
\begin{equation}
 -(i/k)\int_{-\infty}^{\infty}\cos k\xi'\sech^2\xi' d\xi' = -(i/k)
  F(k) 
\end{equation}
where $F(k)\equiv \pi k [2\sinh(\pi k/2)]^{-1}$ which decays rapidly
with large $k$. The $k$ integration in Eq.(A.1) is
\begin{eqnarray}
 {\rm Re}(-i)\int_{-\infty}^{\infty} &dk&e^{ik\xi}\big[\, ik
-\tanh\xi\,\big] k^{-1} F(k)  
\nonumber \\
&=&\int_{-\infty}^{\infty}\cos k\xi \,F(k)\, dk \nonumber \\
 &\hspace{0.3in}-&(\pi/2)\tanh\xi\int_{-\infty}^{\infty} dk\, 
 \sin k\xi\,[\sinh(\pi k/2)]^{-1}\nonumber \\
 & = &  (2/\pi)\bigg( \sech^2\xi - \tanh^2\xi\bigg)\nonumber\\
 & = &  \frac{4}{\pi}\sech^2 \xi  
\end{eqnarray} 
where we have dropped the irrelevant constant $(2/\pi)$. 
The time integral in Eq.(A.1) is $f(t)$ where
\begin{eqnarray}
f(t)&\equiv&\int_0^{\infty}
 dt'\sin\Omega(k)(t-t')e^{-\beta(t-t')}f_1(t') \nonumber \\
  &= & (\epsilon_1/2)\cos\omega
t\bigg\{\frac{\Omega(k)-\omega}{\beta^2+(\Omega(k)-\omega)^2}
 \nonumber \\
& &\hspace{0.8in}+\frac{\Omega(k)+\omega}{\beta^2 + (\Omega(k)+\omega)^2}\bigg\}
 \nonumber \\
&+&(\epsilon_2/2)\cos(2\omega t+\theta)
 \bigg\{\frac{\Omega(k)-2\omega}{\beta^2+(\Omega(k)-2\omega)^2}\nonumber
 \\
& &\hspace{0.8in}+\frac{\Omega(k)+2\omega}{\beta^2 + (\Omega(k)+2\omega)^2}\bigg\}
\end{eqnarray}

In this paper we only consider values of $\omega$ which are
appreciabely less than one. While $\Omega(k) = (1+k^2)^{1/2}$ where in
our units the lower band edge has the value one. Consequently with the
$\omega$'s in this paper as in Ref.[1] \nocite{Flach02}, there is
essentially no radiationof SG phonons generated by the ac driver but
only a dressing of the soliton that is localized on the soliton. The
presence of the SG phonons would not qualitatively alter the symmetry
breaking but for the ac driver frequencies used in this paper the SG
phonons would not be observable because they would occur only in very
high orders of perturbation theory. Since $F(k)$ decreases rapidly with
increasing $k$ and $\omega \ll 1$ we can treat $f(t)$ as effectively
independent of $k$ and equal to

\begin{eqnarray}
 f(t) &=& (\epsilon_1/2)\cos\omega
 t\biggl\{\frac{1-\omega}{\beta^2+(1-\omega)^2} +
 \frac{1+\omega}{\beta^2+(1+\omega)^2}\bigg\} \nonumber \\
 &+ & (\epsilon_2/2)\cos(2\omega t+\theta)
 \biggl\{\frac{1-2\omega}{\beta^2+(1-2\omega)^2} \nonumber \\
& & \hspace{1.3in}+\frac{1+2\omega}{\beta^2+(1+2\omega)^2}\bigg\} \label{ap7}
\end{eqnarray}
Finally we have
\begin{equation}
 \chi = (4/\pi)\, f(t)\,\sech^2\xi
\end{equation}
with $f(t)$ given by Eq.(A.7). 



\end{document}